\documentclass[reprint,superscriptaddress,amsmath,amssymb,aip,jcp]{revtex4-1}

\usepackage{graphicx} 

\draft 

\begin{document}

\title{Thermodynamics of switching in multistable non-equilibrium systems}

\author{Jacob Cook}
\author{Robert G. Endres}
\email[E-mail: ]{r.endres@imperial.ac.uk}
\affiliation{Department of Life Sciences, Imperial College, London SW7 2AZ, United Kingdom}
\affiliation{Centre for Integrative Systems Biology and Bioinformatics, Imperial College, London SW7 2AZ, United Kingdom}

\date{\today}

\begin{abstract}
Multistable non-equilibrium systems are abundant outcomes of nonlinear dynamics with feedback but still relatively little is known about what determines the stability of the steady states and their switching rates in terms of entropy and entropy production.
Here, we will link fluctuation theorems for the entropy production along trajectories with the action obtainable from the Freidlin--Wentzell theorem to elucidate the thermodynamics of switching between states in the large volume limit of multistable systems.
We find that the entropy production at steady state plays no role, but the entropy production during switching is key.
Steady-state entropy and diffusive noise strength can be neglected in this limit.
The relevance to biology, ecological, and climate models is apparent.
\end{abstract}

\maketitle

\section{Introduction}
When Niels Bohr and Erwin Schr\"odinger asked decades ago whether new physical principles are needed to explain living systems, the answer seemed ``no'' \cite{bohr33,schroedinger44}.
More recently, however, the field of stochastic thermodynamics with its temporal violations of macroscopic thermodynamic laws at the microscopic scale have provided a new physical perspective on life.
Most remarkable corner stones of far-from-equilibrium thermodynamics are the fluctuation theorems and Seifert's thermodynamic uncertainty relation, stressing the important role of entropy production \cite{searles99,evans02,seifert12,barato15}.
At equilibrium, detailed balance prohibits any entropy production on average, but far from equilibrium such entropy production is a characteristic feature \cite{battle16} and reflects the flow of time \cite{roldan15}.

The fluctuation theorem by Evans and Searles allows the exact calculation of the entropy production along a trajectory from the time-forward and time-reversed path (corresponding to a movie played backwards), where paths can be calculated from e.g.\ Gillespie simulations of the underlying chemical master equation \cite{evans02}.
However, due to its intrinsic connection with the time-reversed path, the theorem cannot be used to calculate the probability of a path simply from its entropy production.
The situation is different when using the least-action principle, which allows the prediction of the most likely path between two points in a stochastic system from minimising the action (integral over the Lagrangian) \cite{assaf17,arnold00}.
This is often done with a Langevin approximation of the master equation, such as using stochastic differential equations incorporating noise terms \cite{perez16,cruz18}.
However, now the link to thermodynamics is less clear as the role of the entropy production is obscured by the action functional.

In this paper, we combine the fluctuation theorem and least-action principle to address the stability of steady states in non-equilibrium systems.
In particular, we will elucidate the roles of steady-state entropy and fluctuations, as well as steady-state and path entropy production in state switching.
For this purpose, we use two different low-dimensional minimal models shown in Fig.~\ref{fig:fig1}, the Schl\"ogl \cite{schloegl72,vellela09,endres15} and the toggle switch \cite{cherry00} models, with fixed external species concentrations to ensure non-equilibrium behaviour.

\section{Merging two approaches for state switching}
\begin{figure*}[!tbhp]
\includegraphics[width=0.75\textwidth]{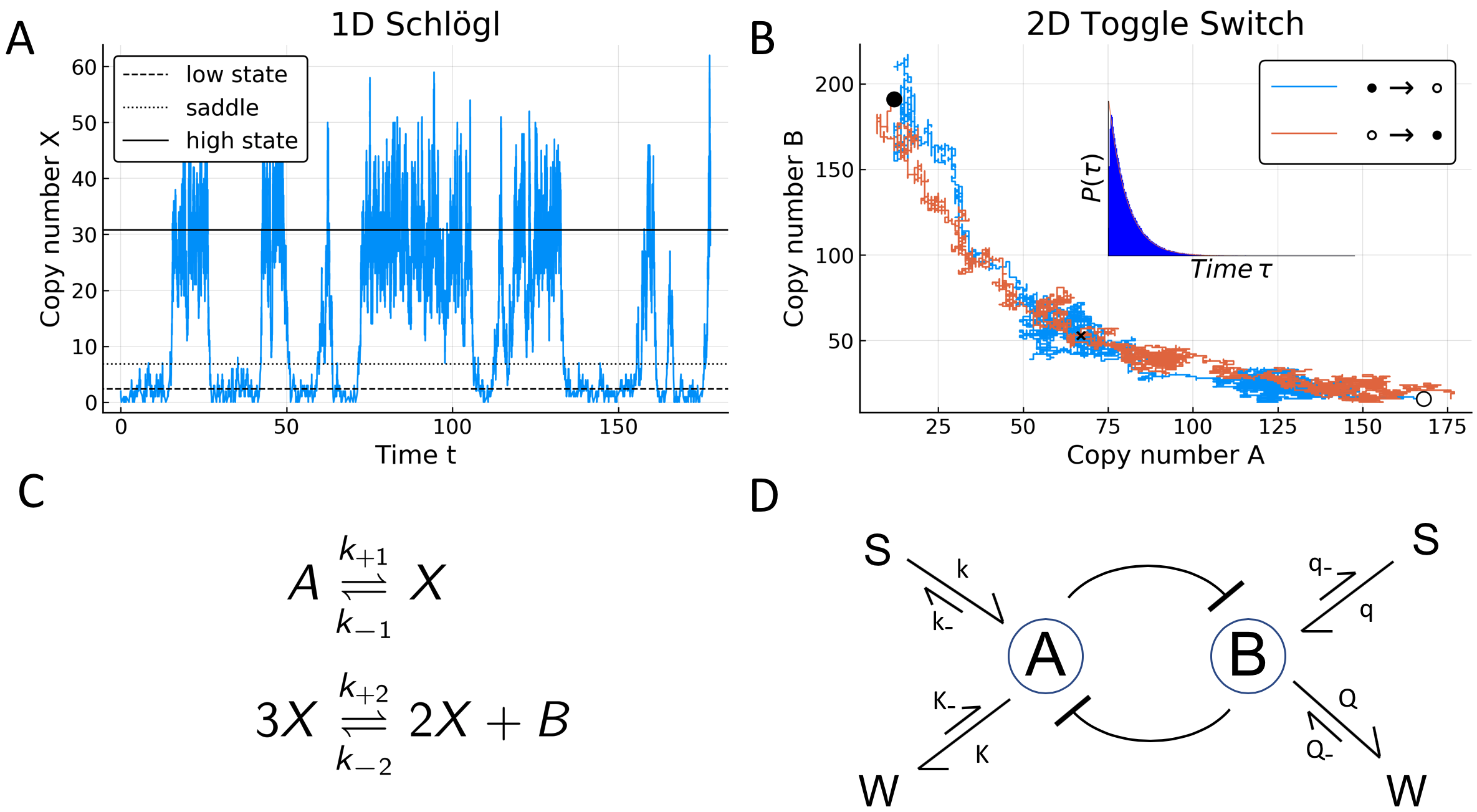}
\caption{{\bf Overview of models.} (a) Example simulation of Schl\"ogl model, displaying switching between high and low $X$ (copy number) states with $A$ and $B$ reservoir species held constant.
(b) Example switching paths between $A$ and $B$ states for toggle switch model. (Inset) Waiting time distribution for A to B switching, with a  clear exponential distribution beyond a small initial time.
(c) Symbolic chemical equations describing the Schl\"ogl model.
(d) Schematic illustration of 4 species toggle switch model, consisting of two mutually repressing chemical species ($A$,$B$) in addition to substrate (S) and waste (W) species. As the mutual repression between the two chemical species occurs via dimer binding the two rates relevant to this ($r$ and $f$) are not included in the schematic.
The parameter values used to generate the example paths can be found in the supplementary material.}
\label{fig:fig1}
\end{figure*}

To investigate the thermodynamics of state switching we shall study non-equilibrium bistable systems with macrostates denoted $A$ and $B$, where both macrostates correspond to sets of microstates in the discrete space of molecule numbers \textbf{X}, which is a vector for multiple chemical species.
The assumption is made that no significant amount of time is spent outside these macrostates.
In the large volume limit, the process of switching between states can be assumed to be a Poisson process (with exponentially distributed waiting times, see Fig.~\ref{fig:fig1}B inset).
Thus, $\langle\tau_A\rangle=k_{A{\rightarrow}B}\int_0^\infty t\exp(-k_{A{\rightarrow}B}t) dt=k_{A{\rightarrow}B}^{-1}$ where $k_{A{\rightarrow}B}$ is the switching rate from $A$ to $B$, and similarly for the $B$ state.
These switching rates are coarse-grained and will in general have contributions from a number of microscopic paths.
The occupation probability of the $A$ state is then given by $p_A=\langle\tau_A\rangle/(\langle\tau_A\rangle+\langle\tau_B\rangle) = 1/(1+k_{A{\rightarrow}B}/k_{B{\rightarrow}A})$.
Hence, such a two-state system is completely described by the ratio of the switching rates.
How do we calculate these for actual molecular systems?
While switching in equilibrium systems is often determined by Kramer's formula, where the height of the energy barrier matters, the treatment of non-equilibrium systems requires entire paths\cite{wang15,feng14}.

\subsection{Microscopic fluctuation theorems}
Using a microscopic description for non-equilibrium systems, the dynamics can be described by a path $\Gamma$, e.g.\ ${\bf X}_0$, ${\bf X}_1$, ${\bf X}_2$,{\dots}, ${\bf X}_N$, obtainable from simulations of the chemical master equation \cite{gaspard04}.
The master equation describes the probability evolution of the microscopic stochastic process and can be formulated as
\begin{equation}
\begin{split}
    \frac{d}{dt}P(\mathbf{X};t) = \sum_{i=1}^{N_\rho}&[W_{\rho_i}(\mathbf{X}-\boldsymbol{\nu}_{i}|\mathbf{X})P(\mathbf{X}-\boldsymbol{\nu}_{i};t)\\
    &- W_{\overline{\rho}_i}(\mathbf{X}|\mathbf{X}-\boldsymbol{\nu}_{i})P(\mathbf{X};t)],
\end{split}
\end{equation}
where $P(\mathbf{X};t)$ is the probability of being in microstate $\mathbf{X}$ at time $t$ and $N_\rho$ is the total number of reactions (including reverse reactions) with $\overline{\rho}_i$ the corresponding reverse reaction to reaction $\rho_i$.
In particular, $W_{\rho_c}(\mathbf{X}_a|\mathbf{X}_b)$ is the transition rate from state $a$ to state $b$ by reaction $\rho_c$, and  $\boldsymbol{\nu}_{i}$ is the change in molecular copy numbers caused by reaction $\rho_i$.
For every time-forward path, there also exists a time-reversed path $\overline{\Gamma}$, ${\bf X}_N$, ${\bf X}_{N-1}$,{\dots}, ${\bf X}_0$.
The probability of observing a particular path, e.g.\ the above time-forward path, is given by the path probability $P_{\Gamma}=\mathcal{N}P({\bf X}_0)\prod^N_{i=1}P(\tau_{i-1,i})W_{\rho_{i}}({\bf X}_{i-1}|{\bf X}_i)$ assuming a continuous-time memory-less Markov process with transition rates $W_{\rho_{i+1}}({\bf X}_i|{\bf X}_{i+1})$.
Specifically, $\rho_i$ is the reaction that causes step $i$, $P(\tau_{i,i+1})$ is the probability that there is no reaction in the time interval between $\tau_{i}$ and $\tau_{i+1}$, and $\mathcal{N}$ is a normalisation factor, ensuring $\sum_{\Gamma}P_{\Gamma} = 1$.
As we are considering a non-equilibrium steady state (NESS) probability distribution, the total entropy production along the path is given by the steady-state fluctuation theorem (FT) \cite{seifert05,malek17} as
\begin{equation}
\Delta S_\Gamma = \ln\left(\frac{P_{\Gamma}}{P_{\overline{\Gamma}}}\right) = \ln\left(\frac{P({\bf X}_0)}{P({\bf X}_N)}\right)+\ln\left(\frac{W_{\Gamma}}{W_{\overline{\Gamma}}}\right),
\label{eq:FT}
\end{equation}
with $W_{\Gamma} = W_{\rho_{1}}({\bf X}_0|{\bf X}_1)\dots W_{\rho_{N}}({\bf X}_{N-1}|{\bf X}_N)$ and $W_{\overline{\Gamma}} = W_{\overline{\rho}_{N}}({\bf X}_N|{\bf X}_{N-1})\dots W_{\overline{\rho}_{1}}({\bf X}_{1}|{\bf X}_0)$ (for details see supplementary material).
The entropy considered here is defined as the Shannon entropy of the probability distribution of states $\mathbf{X}$, i.e.\
\begin{equation}
    S(t) = -\sum_{\mathbf{X}} P(\mathbf{X};t)\ln P(\mathbf{X};t)
\end{equation}
in units where Boltzmann's constant $k_B = 1$.
For rare switching, valid for large volume $\Omega$, the steady-state value of $S$ has contributions from $S_A$ with $\mathbf{X}{\in}A$ and $S_B$ with $\mathbf{X}{\in}B$.

Restricting our consideration to paths that start within macrostate $A$ and end within $B$, we can calculate the ensemble-averaged total entropy production $\Delta S_{A{\rightarrow}B}=\sum_{\Gamma|A{\rightarrow}B} P_\Gamma\Delta S_\Gamma=\langle\ln(P({\bf X}_0)/P({\bf X}_N))\rangle_{A{\rightarrow}B}+\langle\ln(W_{\Gamma}/W_{\overline{\Gamma}})\rangle_{A{\rightarrow}B}$, where the first term on the right-hand side (RHS) corresponds to the average change in entropy of the system between start ${\bf X}_0{\in}A$ and end ${\bf X}_N{\in}B$ (note $N$, ${\bf X}_0$ and ${\bf X}_N$ can vary for different paths).
Further, the second term on the RHS corresponds to the flow of entropy from system to medium (here chemical reservoirs), which is termed entropy production in the medium \cite{seifert12}.
In the limit of small fluctuations (large $\Omega$, or small noise approximation), the first term becomes negligibly small and the entropy produced is given by the second term only, e.g.\ $\Delta S_{A{\rightarrow}B} \approx \langle \ln{\left(W_{\Gamma}/W_{\overline{\Gamma}}\right)}\rangle_{A{\rightarrow}B} \approx \ln\left(k_{A{\rightarrow}B}/k_{\overline{A{\rightarrow}B}}\right)$.
This expression is of limited use in determining state switching rates as $k_{A{\rightarrow}B}$ is intrinsically linked to $k_{\overline{A{\rightarrow}B}}$, i.e.\ the switching rate associated with time-reversal of the paths from $A$ to $B$.
Also note that in general due to the dissipative dynamics the ensemble of time-reversed paths does not correspond to the ensemble of reverse-switching paths, and hence $k_{\overline{A{\rightarrow}B}} \neq k_{B{\rightarrow}A}$ \cite{feng14}.
In order to uniquely determine switching rates we need to utilize a macroscopic formalism.

\subsection{Macroscopic Langevin coarse-graining}
In the macroscopic limit, we can make a continuum approximation of the master equation with the chemical Fokker-Planck equation (FPE), i.e.\ $X_i = x_i\Omega$ for $i=1,...,K$ a $K$-dimensional chemical system, with concentrations $x_i$ and volume $\Omega$.
The chemical Fokker-Planck equation can be derived following Gillespie \cite{gillespie00} as
\begin{equation}
\begin{split}
    \frac{\partial}{\partial t}P(\mathbf{x};t) = &-\sum^{N}_{i=1}\frac{\partial}{\partial{x_i}}\left[f_i(\mathbf{x})P(\mathbf{x};t)\right]\\
    &+ \frac{1}{2} \sum^{N}_{i,j=1}\frac{\partial^2}{\partial{x_i}\partial{x_j}}\left[D_{ij}(\mathbf{x})P(\mathbf{x};t)\right],
\end{split}
\label{eq:CFPE}
\end{equation}
where the deterministic force $\boldsymbol{f}(\mathbf{x})$ is given by
\begin{equation}
    f_i(\mathbf{x}) = \sum^{N_\rho}_{j=1} \nu_{ji} w_{\rho_j}(\mathbf{x})
\end{equation}
and the diffusion matrix $\boldsymbol{D}(\mathbf{x})$ is defined as
\begin{equation}
    D_{ij}(\mathbf{x}) = \sum^{N_\rho}_{k=1}\nu_{ki}\nu_{kj}w_{\rho_k}(\mathbf{x}).
    \label{eq:Diff}
\end{equation}
In the above definitions, $\nu_{ji}$ is the change in molecular copy number of species $i$ that a single occurrence of reaction $\rho_j$ causes, and $w_{\rho_i}(\mathbf{x})$ is the rate (in concentration units) of reaction $\rho_i$ for concentration $\mathbf{x}$, which is the continuum limit of $W_{\rho_i}(\mathbf{X}-\boldsymbol{\nu}_i|\mathbf{X})/\Omega$.
This Fokker-Planck equation is generally a reasonable approximation for large (but finite) values of $\Omega\,\,$ \cite{gillespie00}, particularly near thermodynamic equilibrium \cite{grima11,hanggi84}.
However, while the accurate prediction of switching rates is difficult, the characterization of relative stability is easier.
The chemical Langevin equation corresponds to this Fokker-Planck equation and can be expressed as $\dot x_i=f_i({\bf x})+g_{ij}({\bf x})\xi_j(t)$ with $\xi_j(t)$ uncorrelated white Gaussian noises of zero mean and autocorrelation $\langle\xi_i(t)\xi_j(t')\rangle = \Omega^{-1}\delta_{ij}\delta(t-t')$.
The deterministic force in direction $i$ is given by $f_i$, and $g_{ij}$ determines the propagation of noise from direction $j$ to $i$.
Here, and throughout the paper we adopt the convention that repeated indices (in this case $j$) are summed over.
For the models considered in this paper $g_{ij}$ is always diagonal, representing multiplicative noise.
As Eq.~\ref{eq:CFPE} is an It\^o equation, we treat $g_{ij}$ via It\^o integration \cite{mannella12}.
It is worth noting that our definition of entropy is essentially unchanged as we move to the continuum limit, though it is now the Shannon entropy of a continuous probability distribution.

Paths from this Langevin equation can be treated via path integral methods \cite{chernyak06}, which can be done more simply in our case.
When the probability of escape from a macrostate is sufficiently low, the stochastic transition will be expected to concentrate along a single path $\bf x^*$, with paths significantly diverging having probabilities so low (for large $\Omega$) as to have negligible impact on overall escape probability \cite{touchette09}.
The Wentzel–-Kramers–-Brillouin (WKB) approximation, the classical analogue of the quantum mechanical WKB \cite{assaf17}, can then be used to obtain the conditional probability of this path as  $P_{A{\rightarrow}B}\sim\exp(-\Omega \mathcal{A}[{\bf x^*}])$, where $P_{A{\rightarrow}B}$ is the conditional probability of reaching macrostate $B$ from the initial macrostate $A$, and $\mathcal{A}$ is the action, as derived in \cite{ventsel70}.
The path $\bf x^*$ will thus minimize the action $\mathcal{A}[{\bf x^*}] = \min \mathcal{A}[{\bf x}] = \mathcal{A}_{A{\rightarrow}B}$.
This path has a fixed start and end point at the minima of macrostates $A$ and $B$, respectively.
(These points can be reached from any point in their respective macrostates by a zero action path.
Thus, the minimum action path $\mathbf{x}^*$ can represent the general path between macrostates.)
The relevant action for a path with duration $\tau$ is the Freidlin--Wentzell (FW) action \cite{ventsel70}
\begin{equation}
\mathcal{A}[{\bf x}]=\frac{1}{2}\int_0^\tau(\dot{x}-f)_iD_{ij}^{-1}(\dot{x}-f)_j\,dt,\label{eq:LDT}
\end{equation}
with the diffusion matrix given by $D_{ij}=g_{ik}g_{kj}$.
Any choice of components $g_{ij}$ is acceptable provided that they result in a diffusion matrix that matches Eq.~\ref{eq:Diff}.
As we are considering It\^o integration this action possesses one additional term, which can be neglected in the large $\Omega$ limit \cite{touchette09}.
(If we instead use Stratonovich calculus, three additional terms would appear, which also disappear in the large $\Omega$ limit \cite{arnold00}).
Away from this limit, switching paths no longer pass through the same saddle point as their reverse switching paths \cite{feng14}.
All of our results will pertain to the large $\Omega$ limit.
In this limit, the mean-first passage time (MFPT) is given by $T_{A{\rightarrow}B} \sim P^{-1}_{A{\rightarrow}B}=Q_{A{\rightarrow}B}\exp(\Omega \mathcal{A}_{A{\rightarrow}B})$ or $\ln(T_{A{\rightarrow}B}) = \ln(Q_{A{\rightarrow}B}) + \Omega \mathcal{A}_{A{\rightarrow}B}$, so that as $\Omega$ grows the contribution from the prefactor $Q_{A{\rightarrow}B}$ becomes less important and the second term on the RHS describes the MFPT to logarithmic precision \cite{cruz18,bouchet16}.
This approach is often more accurate than treatment based on the FPE obtained by van-Kampen expansion of the master equation \cite{assaf17}.
While our expression for the MFPT applies in the large deviations limit, alternative forms can be obtained outside of it \cite{fiasconaro05}.

An expression for the entropy production based on the time-reversal of Langevin paths can be obtained by noting that the probability of the most probable switching path $A{\rightarrow}B$ is given by $P_{A{\rightarrow}B} = \exp(-\Omega\mathcal{A}_{A{\rightarrow}B})/Q_{A{\rightarrow}B}$.
The probability of the corresponding time-reversed path $\overline{A{\rightarrow}B}$ is then found to be $P_{\overline{A{\rightarrow}B}} = \exp(-\Omega\mathcal{A}_{\overline{A{\rightarrow}B}})/Q_{A{\rightarrow}B}$, where $\mathcal{A}_{\overline{A{\rightarrow}B}}$ is the action of the time-reversed path.
As factor $Q_{A{\rightarrow}B}$ has not changed, this probability is not in general equal to $P_{B{\rightarrow}A}$, which contains factor $Q_{B{\rightarrow}A}$.
Combining the above two expressions as in Eq.~\ref{eq:FT} generates an expression for the entropy production for Langevin paths as $\Delta S^{L}_{A{\rightarrow}B} = \Omega\left(\mathcal{A}_{\overline{A{\rightarrow}B}} - \mathcal{A}_{A{\rightarrow}B}\right)$ \cite{endres17}.
Substituting Eq.~\ref{eq:LDT} into this relation leads to 
\begin{equation}
\Delta S^{L}_{A{\rightarrow}B} = 2\Omega\int_0^\tau \dot x_i D_{ij}^{-1} f_j\,dt.
\label{eq:Lent}
\end{equation}
Due to our use of a Langevin equation to describe the paths, the above expression is a coarse-grained (apparent) entropy production \cite{mehl12}, which disappears along with the action at steady steady ($\dot x_i=f_i({\bf x})=0$ in the large $\Omega$ limit).
This suggests that only entropy production along the path matters, and that the Langevin formalism within the steady state is equivalent to a quasi-equilibrium.
In the supplementary material we decompose this coarse-grained entropy production into approximate entropy production (EP) and flow (EF) terms, and find that our EP term does generally correlate with the entropy production as calculated from the master equation.
Hence, despite the obvious difference between the entropy productions from the master equation (Eq.~\ref{eq:FT}) and the Langevin paths (Eq.~\ref{eq:Lent}), the latter has predictive power.

We now want to investigate the special case of one-dimensional systems.
To do so we shall use a relation that is valid at every point along an action minimising path \cite{neu18}
\begin{equation}
    \sum_{ij}\dot{x}_iD^{-1}_{ij}\dot{x}_j = \sum_{ij}f_iD^{-1}_{ij}f_j,
   \label{eq:PK}
\end{equation}
which can be interpreted as every point along the minimising path having equal kinetic and potential energy (defined below).
In the one-dimensional case this relation simplifies further to
\begin{equation}
\begin{split}
    \dot{x}_1D^{-1}_{11}\dot{x}_1&=f_1D^{-1}_{11}f_1\\
    |\dot{x}_1|&=|f_1|,
\end{split}
\end{equation}
where $||$ indicates that the magnitude has been taken.
This means that both paths have the same speeds along their lengths as they necessarily pass through the same points (in 1D).
Thus, for one dimensional systems the path $B{\rightarrow}A$ is identical to the time-reversed path $\overline{A{\rightarrow}B}$.
It then follows that for this case $k_{\overline{A{\rightarrow}B}} = k_{B{\rightarrow}A}$, and so the ratio of switching rates are determined solely by the entropy production as
\begin{equation}
    k_{A{\rightarrow}B}/k_{B{\rightarrow}A}=\exp(\Delta S^L_{A{\rightarrow}B}).
    \label{eq:schFT}
\end{equation}
When considering the broader class of multi-dimensional systems such simple relations no longer apply and we instead consider numerical approaches to explicit minimal models.
This will allow us to investigate how the entropy production varies along the path, and whether diffusion strength and steady-state entropies matter.

\section{Schl\"ogl and toggle switch models}
\begin{figure*}[!tbhp]
    \centering
    \includegraphics[width=\textwidth]{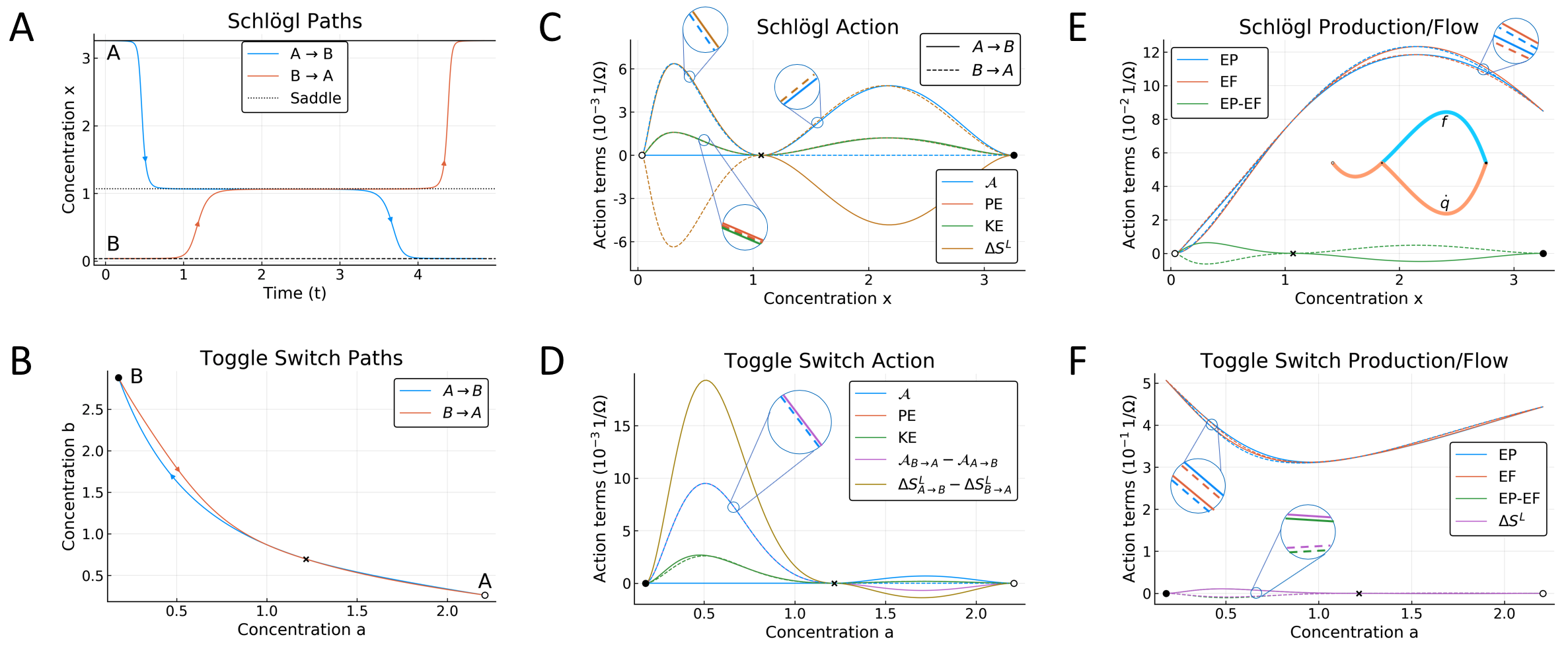}
    \caption{{\bf Dependence of actions on path entropy productions.} (a) Minimum action paths for Schl\"ogl model; aside direction, the only difference between paths is the amount of time spent at the fixed points where there is no contribution to the action. (b) Minimum action paths for toggle switch model, now showing clear differences.
    (c)  Action ($\mathcal{A}$), entropy production ($\Delta S^L$), kinetic (KE) and potential energy (PE) of Schl\"ogl paths with KE and PE defined in Eq.~\ref{eq:consv}, clearly showing that the entropy production of one path is the opposite of the other. In this and the remaining panels a solid line corresponds to the path from $A{\rightarrow}B$ and a dashed line to $B{\rightarrow}A$. Magnifications are meant to clarify line styles.
    (d) Action, differences in action and entropy production, KE and PE along the toggle switch paths. The difference in action is proportional to the difference in entropy produced along the path (this linear relationship is further discussed below).
    (e) Entropy production (EP) and entropy flow (EF) terms along the Schl\"ogl paths. The difference (EP-EF) is equal to the entropy production of time-reversed of Langevin paths. (Inset) Plot showing how $f$ and $\dot{x}$ vary along an exemplar minimum action path. (f) Equivalent plot for the toggle switch paths. The parameter values used to generate the paths can be found in the supplementary material.}
    \label{fig:fig2}
\end{figure*}

The two models considered are the Schl\"ogl and toggle switch models (Fig.~\ref{fig:fig1}).
For both models concentration-constraints are used to make the models non-equilibrium.
The Schl\"ogl model is a simple one-dimensional model that exhibits bistability \cite{schloegl72,vellela09}.
It involves three chemical species $A$, $B$ and $X$, which obey the chemical dynamics shown in Fig.~\ref{fig:fig1}C.
Holding the concentration of external species fixed (i.e. constant $a$ and $b$), this model can be reexpressed in terms of the dynamics of concentration $x$ as
\begin{equation}
    \frac{dx}{dt} = k_{-2}bx^2 - k_{+2}x^3 - k_{-1}x + k_{+1}a + g_{x}(x)\xi(t),
\end{equation}
where
\begin{equation}
    g_x(x) = \sqrt{k_{-2}bx^2 + k_{+2}x^3 + k_{-1}x + k_{+1}a}
\end{equation}
and $(k_{\pm 1},k_{\pm 2})$ are rate constants.
See the supplementary material for further details of this model.

The toggle switch model is a two-dimensional model that describes the dynamics of a simple bistable genetic switch \cite{cherry00} (see Fig.~\ref{fig:fig1}D).
We consider the adiabatic limit where the fraction of genes active is completely determined by the concentration of the corresponding repressing protein (see supplementary material for details).
This means that the model can be expressed in terms of change of protein concentrations $a$ and $b$ as
\begin{align}
\frac{da}{dt} &= \frac{kr}{r + fb^2} - k_{-}a - Ka + K_{-} + g_a(a,b)\xi_a(t)\\
\frac{db}{dt} &= \frac{qr}{r + fa^2} - q_{-}b - Qb + Q_{-} + g_b(a,b)\xi_b(t),
\end{align}
where
\begin{equation}
    g_a(a,b) = \sqrt{\frac{kr}{r + fb^2} + k_{-}a + Ka + K_{-}},
\end{equation}
and 
\begin{equation}
    g_b(a,b) = \sqrt{\frac{qr}{r + fa^2} + q_{-}b + Qb + Q_{-}}.
\end{equation}
Further, $r$ is the gene activation rate and $f$ is rate constant for gene repression by protein dimers, the other eight rate constants describe the rates of protein production or degradation.
Here, the constant external concentrations $s$ and $w$ have been absorbed into the relevant rate constants.

In order to study the switching between states for these models, we need to find paths that minimise the action (Eq.~\ref{eq:LDT}).
These paths are usually determined by use of a quasi-Newton method (e.g. the L-BFGS algorithm) to find the minimising path for a particular duration $\tau$, with a gradient descent method used to find the value of $\tau$ that results in the lowest minimum \cite{perez16}.
However, in order to save computational time we made use of the faster (but more complicated) geometric minimum action method \cite{heymann08,heymann08-2,neu18}.
These paths were then be used in Eq.~\ref{eq:LDT} and Eq.~\ref{eq:Lent} in order to generate the corresponding minimum actions and path entropy productions.

\begin{figure*}[!tbhp]
    \centering
    \includegraphics[width=0.75\textwidth]{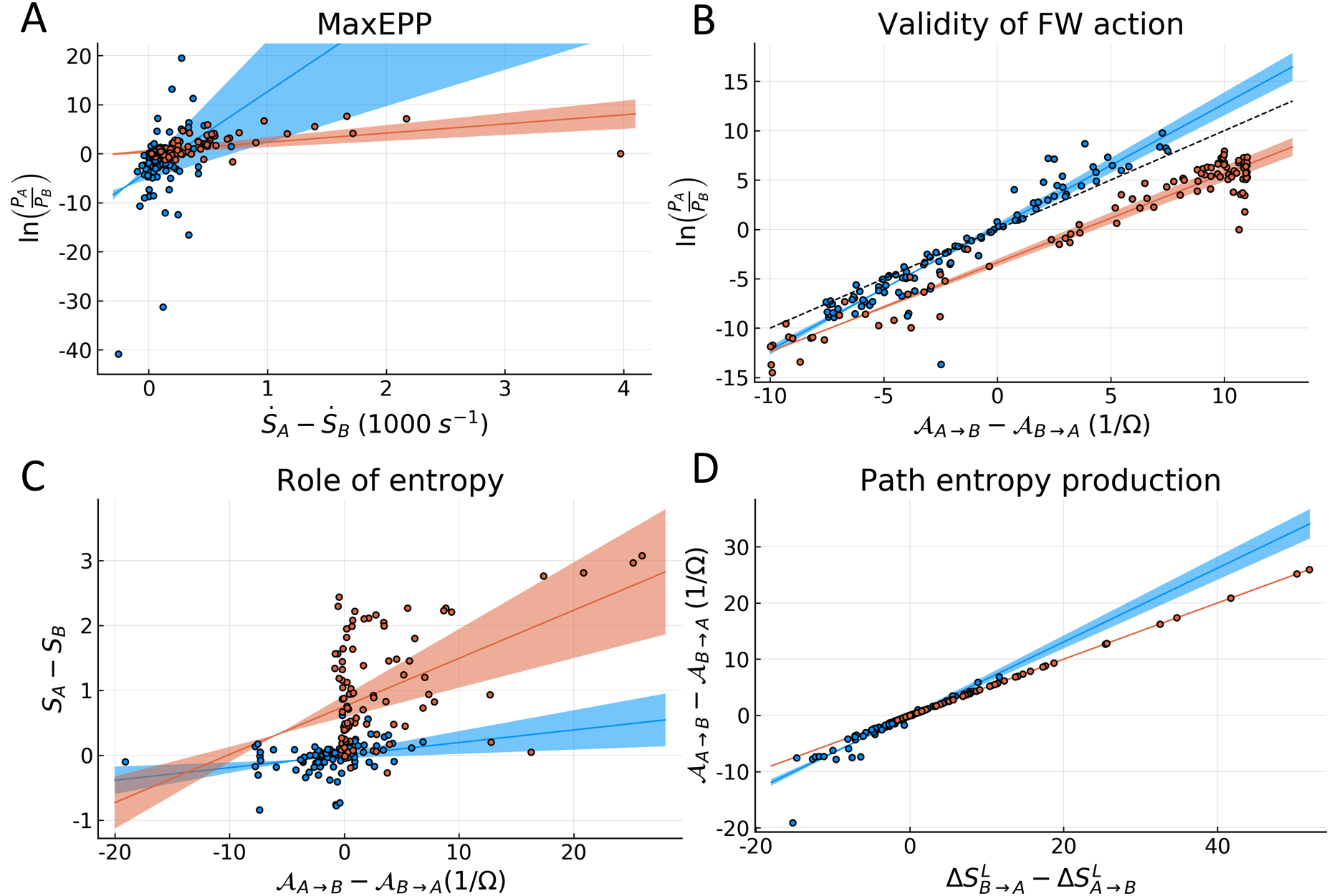}
    \caption{{\bf Comparison of states and switching paths.} In all panels, red and blue dots denote specific parameterizations of Schl\"ogl and toggle switch models, respectively. In each plot the lines and shaded regions indicate best fits and $95\%$ confidence intervals for the particular data sets, respectively. (a) Comparison of log ratios of occupation probabilities vs difference in entropy productions at steady states from Gillespie simulations. The Schl\"ogl data has a Pearson correlation of $0.5172$ and the toggle switch has $0.3476$.
    (b) Log ratio of occupation probabilities obtained from Gillespie simulation vs the difference in minimum action. The dashed line indicates a perfect correspondence. The toggle switch and Schl\"ogl data have correlations of $0.9515$ and $0.9738$, respectively.
    The results shown are coarsely discretized due to the low $\Omega$ used to save computational time.
    The discretization will effect the Schl\"ogl $B$ state disproportionately as it is formed of significantly fewer microstates than $A$.
    This represents a potential explanation for the downwards shift of the Schl\"ogl data.
    (c) Comparison of difference in entropy of steady states vs difference in action. Entropies were found by Gillespie simulation with $\Omega=1$. Both sets of data show weak correlations of $0.2892$ and $0.4622$ for the Schl\"ogl and toggle switch models, respectively.
    (d) Difference in action vs difference in entropy produced along paths. Both models display a strong linear relationship, with correlations of $0.9445$ and $1.0000$ for the toggle switch and Schl\"ogl models, respectively.}
    \label{fig:fig3}
\end{figure*}

Figs.~\ref{fig:fig2}A,B show exemplar minimum action paths for switching in the Schl\"ogl and the toggle switch model, respectively.
As the Schl\"ogl model is one-dimensional its time-reversed switching paths correspond to the switching paths for the contrary direction.
Fig.~\ref{fig:fig2}C shows how this leads to equal and opposite entropy productions (orange lines) along the paths.
For the more complicated toggle switch model this simple relation between paths is lost, but Fig.~\ref{fig:fig2}D shows a linear relation between the difference in minimum action (purple line) and the difference in path entropy productions (gold line).
Despite the systems' quasi-equilibrium behavior, non-equilibrium processes still occur.
In order to illustrate this fact we derived approximate entropy production and flow terms (details in supplementary material).
Figs.~\ref{fig:fig2}E,F show plots of the derived entropy production (EP, blue lines) and flow (EF, red lines), demonstrating non-zero contributions at the steady states.

In order to investigate links between occupation probabilities, entropy, and entropy production, 100 random parameters sets were created for each model.
Specifically, the procedure for the construction of the toggle switch parameter sets was as follows.
First, $k$ was randomly drawn from the (continuous) range $[1,100]$ and $K$ from the range $[0.1,10]$.
Then random numbers were drawn from the range $[0.001,0.1]$ to obtain the ratios $q/k$, $k_{-}/k$, $q_{-}/k$, $Q/K$, $K_{-}/K$, and $Q_{-}/Q$.
Finally, $r$ and $f$ were both drawn from the range $[0,10^4]$.
For the Schl\"ogl parameter sets, all four parameters ($k_{-1}$, $k_{+1}a$, $k_{+2}$, $k_{-2}b$) were drawn from the range $[0.1,10]$.
In both cases the constructed sets were accepted if they resulted in multiple non-zero steady states.
State entropies and occupation probabilities were obtained by direct Gillespie simulation of the chemical master equation \cite{gillespie77}.
Each of these simulations was run for the largest computationally feasible value of $\Omega$ in order to minimise the effect of discretization.
Steady-state entropy productions were calculated using the Schnakenberg method (product of flux and reaction affinity) \cite{schnakenberg76}.

Fig.~\ref{fig:fig3}A shows a weak correlation between state occupation probability and steady-state entropy production, which provides some evidence for the maximum entropy production principle (MaxEPP) \cite{endres17}.
This extremal principle proposes that states with higher entropy production are more dynamically stable (subject to other dynamical constraints) \cite{dewar09,lorenz01}.
We then approximate the log ratio of state occupation probabilities via the Freidlin--Wentzell theorem as, $\ln{(p_A/p_B)} \approx \ln{(P_{B{\rightarrow}A}/P_{A{\rightarrow}B})} = \ln{(Q_{B{\rightarrow}A}/Q_{A{\rightarrow}B})} + \Omega\left(\mathcal{A}_{A{\rightarrow}B} - \mathcal{A}_{B{\rightarrow}A}\right)$.
For large $\Omega$ only the second term would be expected to contribute but this limit is difficult to simulate.
Simulated occupation probabilities match well with this approximation (see Fig.~\ref{fig:fig3}B), demonstrating the validity of our use of the FW action.
Fig.~\ref{fig:fig3}C shows a weak correlation between difference in action and difference in state entropy, as expected from equilibrium theory where higher entropy states are more stable.
However, state entropy increases sublinearly with $\Omega$ so for large $\Omega$ it has no effect on the stability.
Fig.~\ref{fig:fig3}D shows a comparison of the difference in action and the difference in path entropy production, showing that the linear relation observed in Fig.~\ref{fig:fig2}C,D holds generally across parameter sets surveyed.
The effect of diffusion strength was found to be minimal (plots of this are therefore provided in the supplementary material).

Our results suggest a limited form of MaxEPP, which applies to the rate of switching between macrostates.
We shall proceed with our derivation by noting that the action can be split into two parts as $\Omega\mathcal{A}_{A{\rightarrow}B} = \mathcal{C}_{A{\rightarrow}B} - \frac{1}{2}\Delta S^{L}_{A{\rightarrow}B}$, where $\mathcal{C}_{A{\rightarrow}B}$ is the conservative action along the path $A{\rightarrow}B$ and $\Delta S^{L}_{A{\rightarrow}B}$ is the Langevin path entropy production (Eq.~\ref{eq:Lent}) \cite{endres17}.
The conservative action can be expressed in a similar form to Eq. \ref{eq:Lent} as
\begin{equation}
\mathcal{C}_{A{\rightarrow}B} = \frac{\Omega}{2}\int^{\tau}_{0}\left(\dot{x}_iD_{ij}^{-1}\dot{x}_j + f_i\,D_{ij}^{-1} f_j\right)\,dt,
\label{eq:consv}
\end{equation}
where the two terms resemble kinetic (KE) and potential energy (PE) contributions, respectively.
By substituting the expanded form of the action into the expression for switching path probability, a reduced form of MaxEPP can be obtained
\begin{equation}
P_{A{\rightarrow}B} = \frac{\exp{\left(\frac{1}{2}\Delta S^{L}_{A{\rightarrow}B} - \mathcal{C}_{A{\rightarrow}B}\right)}}{Q_{A{\rightarrow}B}},
\label{eq:MaxEPP}
\end{equation}
where $P_{A{\rightarrow}B}$ is the probability of the (most probable) switching path along $A{\rightarrow}B$, and $Q_{A{\rightarrow}B}$ is a constant.
This equation shows that there is a trade-off between minimization of the conservative action (i.e.\ fulfilling the equation of motion) and maximization of the path entropy production (i.e.\ being as dissipative as possible).

If the switching path and its contrary path pass through similar regions of space, then they will have similar kinetic energy contributions along their lengths provided that the deterministic force $\mathbf{f}$ does not vary too rapidly.
From Eq.~\ref{eq:PK} it can been seen that at every point on a minimizing path the potential energy contribution is equal to the kinetic energy contribution.
The approximation that $C_{A{\rightarrow}B}\approx C_{B{\rightarrow}A}$ can thus reasonably be made.
This approximation is exact for the 1D Schl\"ogl model as $\overline{A{\rightarrow}B} = B{\rightarrow}A$, and holds for 90\% of the toggle switch parameterizations used.
In contrast, the dissipative (path entropy production) component Eq.~\ref{eq:Lent} depends on cross terms of velocity $\mathbf{\dot{x}}$ and deterministic force $\mathbf{f}$, and as such are not be expected to cancel, leading to
\begin{equation}
     \frac{1}{2}\left(\Delta S^{L}_{B{\rightarrow}A} - \Delta S^{L}_{A{\rightarrow}B}\right)  \approx \Omega\left(\mathcal{A}_{A{\rightarrow}B} - \mathcal{A}_{B{\rightarrow}A}\right)
\label{eq:semian}
\end{equation}
in line with the relationship seen in Fig.~\ref{fig:fig3}D.
Significant divergence from the relation was generally observed in cases where the saddle point occurred at a low copy number compared to the steady states, due to the substantially faster variation of the force.
Further discussion of the above derivation can be found in the supplementary material.

\section{Discussion}
Our primary conclusion is that a MaxEPP for switching paths can be obtained within the Langevin approximation (Eq. \ref{eq:MaxEPP}), extending the rule that ``exergonic reactions occur spontaneously'' to switching in multistable systems.
In a system with a large number of potential macrostates our relation predicts that for sufficiently large volumes switches that produce more entropy will be favoured.
If regions of state space with greater entropy productions also require greater path entropy productions to reach, then this could form a basis for a more extensive maximum entropy production principle.
Our secondary conclusion is that there exists a relationship between the difference in action of minimum action paths and difference in entropy produced along these paths (Eq.~\ref{eq:semian}), valid for all paths that do not pass through regions of rapidly varying force.
We additionally found that steady-state entropies had very little effect on the stability of steady states (see Fig.~\ref{fig:fig3}C).
Finally, we found some evidence to support a broader maximum entropy production principle (see Fig.~\ref{fig:fig3}A).

There has been significant interest in the thermodynamics of the transition between different steady-state probability distributions when controlled by an external parameter \cite{bagci13,hatano01}.
This is fundamentally different to our work, which is about the thermodynamics of switching in a bistable system.
In our two-state systems, we would naively expect a net zero entropy production through switching as the entropy produced by a switch in one direction would be cancelled by the eventual switch back.
Only in cases where the switching path differs from the converse switching path are there net fluxes of probability through the system and thus entropy production.
Consistently, for the Schl\"ogl model we find no net entropy production (i.e.\ $\Delta S^L_{A\rightarrow B} = -\Delta S^L_{B\rightarrow A}$).
In other recent work\cite{ruelle16}, bounds on the ratio of transition rates between states in a bistable system based on relative entropy $\Delta H$ and path entropy production are found.
This ratio is determined as $\pi(B\rightarrow A)/\pi(A\rightarrow B) \geq \exp{[-\Delta S_{A\rightarrow B}]}$, where $\pi(A\rightarrow B)$ is the sum of the rates of switching $A$ to $B$ over all possible switching channels.
With the expectation that a single most probable path will dominate, consideration of the two contrary switching paths will be sufficient.
Combining the main result of their paper with our analytic relation (Eq. \ref{eq:semian}) leads to a bound on the path entropy produced as $\Delta S^{L}_{A\rightarrow B}+\Delta S^{L}_{B\rightarrow A}\geq 0$, which becomes an equality in the limit of time-reversed switching paths (e.g.\ Schl\"ogl).
Every parameter set used in Fig.~\ref{fig:fig3} was found to satisfy this condition.

Beyond the chemical physics literature, related frameworks to ours have been used in evolutionary science \cite{vladar11,mustonen10} where the cumulative fitness flux is maximized (like entropy production) subject to the trade-off that speed of allele change and magnitude of selective forces are minimized (like the conservative action).
Our results therefore suggest, that states in evolutionary systems that require greater cumulative fitness fluxes to reach should be expected to be more stable.
Exploring the application of our theory in ecology, and evolutionary science with systems of multiple stable states will be an interesting way forward \cite{scheffer09}.

\section*{Supplementary material}
See supplementary material for further details of the models used, details of how entropy production was calculated from Gillespie simulations, extended derivations of our relations, details of algorithms used to minimize actions, extended discussion of our coarse-grained entropy production, a plot showing the limited effect of diffusion strength, and tables of parameters for Figs.~\ref{fig:fig1} and \ref{fig:fig2}. 

\begin{acknowledgments}
J.C. and R.G.E. thankfully acknowledge helpful discussions with Shamil Chandaria, Henrik Jensen, Chiu Fan Lee, Gunnar Pruessner, David Schnoerr, Philipp Thomas, and financial support from the NERC CDT in Quantitative and Modelling Skills in Ecology and Evolution (grant No./ NE/P012345/1), and BBSRC grant BB/N00065X/1.
\end{acknowledgments}

\bibliography{main}

\end{document}